\documentclass[prl,twocolumn,%showpacs,
tightenlines,superscriptaddress,preprintnumbers,floatfix]{revtex4}

\usepackage{graphicx}

\newcommand{\beq}{\begin{equation}}
\newcommand{\eeq}{\end{equation}}
\newcommand{\bqa}{\begin{eqnarray}}
\newcommand{\eqa}{\end{eqnarray}}

\preprint{INT PUB 07-26}

\begin{document}

\title{Fluid turbulence and eddy viscosity 
in relativistic heavy-ion collisions}

\author{Paul Romatschke}
\affiliation{Institute for Nuclear Theory, University of Washington,
Box 351550, Seattle WA, 98195, USA}
\date{\today}

\begin{abstract}
The eddy viscosity for a turbulent compressible 
fluid with a relativistic equation of
state is derived. Compressibility allows for sound modes, but the eddy
viscosity in the shear mode is found to be the same as for 
incompressible fluids. For two space dimensions (which is the relevant 
case for the dynamics of relativistic heavy-ion collisions) 
the eddy viscosity in the shear mode is negative, reducing the effective
viscosity below its microscopic value. This could explain
the tiny viscosity found at RHIC. Implications for
the experimentally accessible elliptic flow coefficient 
at the LHC are speculated on.
\end{abstract}

\maketitle

Efforts to understand the bulk physics of the ongoing experimental 
program at the Relativistic Heavy-Ion Collider (RHIC)
have led to an interest in the theory of relativistic fluid dynamics,
both with and without shear viscosity. The curious fact that
ideal fluid dynamics is able to describe the experimentally measured
particle spectra amazingly well 
\cite{Teaney:2001av, Huovinen:2001cy,Kolb:2001qz,Hirano:2002ds,Kolb:2002ve} 
has led to the hypothesis that
the shear viscosity $\eta$, or rather the relevant dimensionless quantity
$\eta/s$ involving the entropy density $s$, has to be extremely small
\cite{Shuryak:2004cy,Heinz:2005zg}.
Indeed, values of $\eta/s\sim 1$ (in natural units $k_B=\hbar=c=1$) obtained by solving Quantum-Chromodynamics (QCD) in the weak coupling
expansion \cite{Arnold:2003zc} were seemingly ruled out \cite{Teaney:2003kp}.
Interest in RHIC spread much beyond nuclear physics 
when it was realized that 
a calculation of $\eta/s$ by string theory methods for a 
thermal gauge theory in the strong coupling limit \cite{Policastro:2001yc} 
gave $\eta/s=\frac{1}{4 \pi}$, about an order of magnitude smaller 
than the weak coupling QCD result. 

While this result had originally been
derived not for QCD but for the ${\mathcal N}=4$ supersymmetric Yang-Mills
theory, it was later conjectured \cite{Kovtun:2004de} 
that all relativistic quantum field
theories at finite temperature and zero chemical potential obey
$\eta/s\ge\frac{1}{4 \pi}$. No laboratory fluid seemed to violate this
bound until recently, when it was shown 
\cite{Romatschke:2007mq} that the experimental data
from top RHIC energies actually favored $\eta/s\sim\frac{1}{8 \pi}$ (another
group argues for even smaller values \cite{Song:2007fn}). While addressing 
the caveats outlined in \cite{Romatschke:2007mq} 
will ultimately allow to decide whether
RHIC violates the bound $\eta/s\ge\frac{1}{4 \pi}$ or not,
it is probably fair to conclude that $\eta/s$ at RHIC is apparently
extremely small (see also \cite{Gavin:2006xd,Lacey:2006bc,Adare:2006nq}). 

The Reynolds number for a relativistic
fluid at temperature $T$ can be estimated as 
${\rm Re}\sim \frac{s}{\eta} T L$, where $L$ is a typical length
scale. For gold collisions at RHIC, taking $L$ to be the radius of a 
gold nucleus $L\sim 6$ fm and the QCD scale as temperature 
($T\sim 200$ MeV) imply
\beq
{\rm Re}_{RHIC}\sim6 \frac{s}{\eta} \sim 48 \pi \gg 1,
\eeq
where the value for $\eta/s$ from \cite{Romatschke:2007mq} was used.
One is thus led to wonder about the possibility and
consequence of fluid turbulence in heavy-ion collisions at RHIC.

The natural starting point are the relativistic fluid dynamic
equations in the presence of shear viscosity
\bqa
(\epsilon+p)D u^\mu&=&\nabla^\mu p-
\Delta^\mu_\alpha D_\beta \Pi^{\alpha \beta}\, ,
\nonumber\\
D \epsilon &=& - (\epsilon+p) \nabla_\mu u^\mu+\frac{1}{2}\Pi^{\mu \nu}
\langle\nabla_\nu u_\mu\rangle\, ,
\label{baseq}
\eqa
where $\epsilon,p$ are the energy density and pressure, $u^\mu$
is the fluid four velocity obeying $u^\mu u_\mu =1$ and 
$\Pi^{\mu \nu}$ is the shear tensor which
to lowest order in gradients \footnote{
To this order, causality is violated at high wavenumbers \cite{oldpap}. 
Here, this does not pose a problem since only the low wavenumber
limit is considered.
%Here we don't care since we work analytically 
%to lowest order, anyways.
} is
$\Pi^{\mu \nu}=\eta \langle\nabla^\mu u^\nu\rangle$.
The remaining symbols in Eq.~(\ref{baseq}) are: $D_\alpha$, the covariant
derivative, $D\equiv u^\mu D_\mu$, $\nabla_\mu \equiv \Delta^\nu_\mu D_\nu$,
$\Delta^{\mu \nu}\equiv g^{\mu \nu}-u^\mu u^\nu$ and
$\langle\nabla^\mu u^\nu\rangle\equiv\nabla^\mu u^\nu+\nabla^\nu u^\mu-\frac{2}{d}\Delta^{\mu \nu}\nabla_\alpha u^\alpha$. The metric signature
is $g^{\mu \nu}=(+,-,-)$ and $(+,-,-,-)$ for $d=2$ and $d=3$ space dimensions, 
respectively. 
The equation of state $p=p(\epsilon)$ 
is used to close the system (\ref{baseq}). In what follows,
an equation of state with a constant speed of sound squared $c_s^2=d p/d\epsilon$ will be
considered.

Eq.~(\ref{baseq}) is much more non-linear than the non-relativistic
Navier-Stokes equation for incompressible fluids, which is
the canonical starting point for turbulence theory.
In order to make progress, it is therefore useful to consider the
approximation where both $\eta/s$ and the fluid three velocity $u^i$ 
are small, $\eta/s\sim u^i\sim {\mathcal O}(o)\ll 1$.
Introducing the Fourier transform for $d+1$ dimensional space-time
\beq
\ln \epsilon (t,{\bf x}) \equiv \int \frac{d \omega d{\bf k}}{(2\pi)^{d+1}}
\exp [-i \omega t + i {\bf k} \cdot {\bf x}] \ln \epsilon(\omega, {\bf k})
\eeq
and similarly for $u^i$, one finds (in flat space) from Eq.~(\ref{baseq}):
\bqa
\left(D_0^{-1}(K)\right)^{ij}
u^j_K 
\!\!&=&\!\! \frac{i}{2} \int_Q u_Q^m u^l_{K-Q} P^{ilm} (K,Q)
+{\mathcal O}(o^3) \label{almostNS}\\
G_0^{-1}(K)\ln \epsilon_K
\!\!&=&\!\! (1\!+\!c_s^2)\int_Q u^i_Q u^j_{K-Q} Q^{ij}(K,Q)
+{\mathcal O}(o^3)\nonumber\\
\left(D_0^{-1}(K)\right)^{ij}\!\!&=&\!\!\left[
(- i \omega +\nu k^2) \delta^{ij}+k^i k^j \left(\nu \frac{d-2}{d}+
i \frac{c_s^2}{\omega}\right)
%\nu \left(\frac{k^i k^j}{3}+k^2 \delta^{ij}\right)
%+i c_s^2 \frac{k^i k^j}{\omega} 
\right]\nonumber\\
G_0^{-1}(K)\!\!&=&\!\!\left[-\omega^2+c_s^2 k^2- \frac{2(d-1)}{d} i \omega \nu k^2\right]\nonumber\\
P^{ilm}(K,Q)\!\!&=&\!\!c_s^2 k^i \delta^{lm} +\delta^{il} \left(q^m-k^m\right)
-\delta^{im}q^l\nonumber\\
Q^{ij}(K,Q)\!\!&=&\!\!
\left(\frac{\omega^2}{2} +\frac{d-1}{d} i \nu \omega k^2\right) \delta^{ij}
-k^i  (k^j-q^j),\nonumber
\eqa
where $K=(\omega,{\bf q})$, and 
$\eta/(\epsilon+p)$ was approximated by its space-time
average, $\nu\equiv\int dt d{\bf x} \eta/(\epsilon+p)$. This approximation 
-- though not strictly justified -- will allow to connect to 
literature on non-relativistic turbulence, where $\nu$ is 
referred to as kinematic viscosity.

Following the Yakhot-Orszag approach to turbulence \cite{YO},
Eq.~(\ref{almostNS}) is replaced by a more general equation
\beq
\left(D_0^{-1}(K)\right)^{ij}
u^j_K 
= f^i(K)+\frac{i}{2} \int_Q u_Q^m u^l_{K-Q} P^{ilm} (K,Q),
\label{leequation}
\eeq
where $f^i$ is a random force used to model turbulent stirring at 
high wavenumbers $k>\Lambda$. Its correlator in $d$ space dimensions
is taken to be Gaussian
\beq
<f^j(K) f^i (K^\prime)>= 2 \alpha k^{-d} (2 \pi)^{d+1} \delta^{ij} 
\delta(K+K^\prime) 
\label{correlator}
\eeq
with strength $\alpha$ and zero mean $<f^i(K)>=0$. Splitting 
$u_K^i=u_K^{i,<} + u_K^{i,>}$ into a low and high wavenumber part
($u_K^{i,<} = u_K^{i} \theta(\Lambda-k)$), the aim is then to
derive an averaged equation for the geometric flow $u^{i,<}$ 
in the presence of the fluctuating $u^{i,>}$,
\bqa
u^{j,<}_K &=& 
\frac{i}{2} D_0^{ij}(K) \theta(\Lambda-k)\int_Q P^{ilm} (K,Q) \times
\nonumber\\
&&\hspace*{1cm}\left(u_Q^{m,<} u^{l,<}_{K-Q} %\right.
%&&\hspace*{4cm}\left.
+ <u_Q^{m,>} u^{l,>}_{K-Q}>\right).
\eqa
To calculate the mean $<u_Q^{m,>} u^{l,>}_{K-Q}>$, one solves
Eq.~(\ref{leequation}) for $u^{>}$ recursively in $u^<$ \cite{YO}.
Using Eq.~(\ref{correlator}) and focusing on the term linear in $u^<$ 
it follows that
\bqa
&<u_Q^{m,>} u^{l,>}_{K-Q}>=4 i \alpha\ u^{a,<}_K q^{-d} D_0^{m c}(Q) D_0^{b c}(-Q)  \times& \nonumber\\
&D_0^{l h}(K\!-\!Q) P^{hab} (K\!-\!Q,-Q) \theta(q\!-\!\Lambda) 
\theta(|{\bf k\!-\!q}|\!-\!\Lambda).&
\eqa
Defining the projectors $A^{ij}_K=\delta^{ij}-B^{ij}_K$ and 
$B^{ij}_K=\frac{k^i k^j}{k^2}$, one can 
decompose the propagator $D_0$ into a shear and a sound mode
(similar to the Kovasznay modes \cite{smitdus}),
\bqa
&\left(D_0^{-1}(K)\right)^{ij} = \alpha_K A^{ij}_K + \beta_K B^{ij}_K&
\\
%A^{ij}_K&=&\delta^{ij}-B^{ij}_K\\
%B^{ij}_K&=&\frac{k^i k^j}{k^2}\\
&\alpha_K=-i \omega+ k^2 \nu,\qquad
\beta_K=-i\omega + i c_s^2 \frac{k^2}{\omega} + \frac{2(d-1)}{d} \nu k^2 &.
\nonumber
\eqa
The inversion of $D_0^{-1}$ is then straightforward and one finds
\bqa
u^{j,<}_K &=& 
 D_0^{ij}(K) 
\left[\theta(\Lambda-k)\frac{i}{2}\int_Q P^{ilm} (K,Q) u_Q^{m,<} u^{l,<}_{K-Q} \right.\nonumber\\
&&\left.\hspace*{4cm}- R^{i a} u^{a,<}_K\right] \label{introR}\\
%\eqa
%\bqa
R^{i a}&=&2 \alpha \int_{Q,>} q^{-d}  P^{ilm} (K,Q) P^{hab} (K-Q,-Q)  \times
\nonumber\\
&&%\hspace*{1cm}
\left[\frac{A^{mb}_Q A^{lh}_{K-Q}}{\alpha_{Q} \alpha_{-Q} \alpha_{K-Q}}
+\frac{A^{mb}_Q B^{lh}_{K-Q}}{\alpha_{Q} \alpha_{-Q} \beta_{K-Q}}
\right.
\nonumber\\
&&\hspace*{1cm}\left.+\frac{B^{mb}_Q A^{lh}_{K-Q}}{\beta_{Q} \beta_{-Q} \alpha_{K-Q}}
+\frac{B^{mb}_Q B^{lh}_{K-Q}}{\beta_{Q} \beta_{-Q} \beta_{K-Q}}
\right]
\eqa
where $\int_{Q,>}\equiv\int_Q \theta(q-\Lambda) \theta(|{\bf k-q}|-\Lambda)$.
 The frequency
integrations are readily performed and in the small wavenumber limit
$\omega,{\bf k} \rightarrow 0$ one finds
\bqa
\int \frac{d q_0}{2\pi} \frac{1}{\alpha_{Q} \alpha_{-Q} \alpha_{K-Q}} 
&=& \frac{1}{2 q^2 \nu^2 (q^2+|{\bf k-q}|^2)}\nonumber\\
\int \frac{d q_0}{2\pi} \frac{1}{\alpha_{Q} \alpha_{-Q} \beta_{K-Q}} 
&=& \frac{1}{2 c_s^2 |{\bf k-q}|^2} + {\mathcal O}(\nu^2)\nonumber\\
\int \frac{d q_0}{2\pi} \frac{1}{\beta_{Q} \beta_{-Q} \alpha_{K-Q}} 
&=& {\mathcal O}(\nu)\nonumber\\
\int \frac{d q_0}{2\pi} \frac{1}{\beta_{Q} \beta_{-Q} \beta_{K-Q}} 
&=& \frac{d^2}{8 (d-1)^2 q^2 \nu^2 (q^2+|{\bf k-q}|^2)},\nonumber
\eqa
where only terms larger than ${\mathcal O}(\nu)\sim {\mathcal O(o)}$ were kept.
Again, $R^{ia}$ can be decomposed as $R^{ia}=r_A A^{ia}_K + r_B B^{ia}_K$.
%For a qualitative understanding on the effect of turbulence it is
%sufficient to evaluate $r_A$ (calculation of $r_B$ is straightforward,
%but a little tedious). 
From Eq.~(\ref{introR}) this implies that
\beq
(D^{ij})^{-1}(K) u_K^{j,<}=
 \theta(\Lambda-k)\frac{i}{2}\int_Q P^{ilm} (K,Q) u_Q^{m,<} u^{l,<}_{K-Q} 
\eeq
where $D^{ij}=(D_0^{ij})^{-1}(K) +R^{ij}$ is 
the propagator involving turbulence corrections,
\beq
(D^{ij})^{-1}=(\alpha_K+r_A) A^{ij}_K+(\beta_K+r_B) B^{ij}_K.
\label{fullprop}
\eeq
To evaluate $r_A$ in the small ${\bf k}$ limit, one 
contracts $R^{ia} A^{ia}_K$ and upon
shifting ${\bf q}\rightarrow {\bf q+\frac{1}{2} k}$
and using the symmetries of the integral
% ${\bf k}\leftrightarrow -{\bf k}$
obtains (see also \cite{YO})
\beq
r_A=\alpha \nu^{-2} k^2 \frac{2 \pi^{d/2}}{(2\pi)^d\Gamma(d/2)} \frac{d^2 -d -4}{8 d (d+2)} \Lambda^{-4} + {\mathcal O} (k^3)
\label{rA}
\eeq 
Interestingly, this result is identical to that for incompressible fluids
\cite{YO} and 
%It seems that 
modifications for compressible
fluids only arise in the sound channel of 
the %hydrodynamic 
propagator (which is absent for incompressible fluids).
The equation for the full propagator Eq.~(\ref{fullprop}) then implies
that
\beq
\alpha_K+r_A=- i \omega + k^2 \nu_{eff},
\eeq
where the effective turbulent viscosity $\nu_{eff}=\nu+\nu_{eddy}$
being the sum of microscopic and eddy viscosity
\beq
\nu_{eddy}=\alpha \nu^{-2}\frac{2 \pi^{d/2}}{(2\pi)^d\Gamma(d/2)} \frac{d^2 -d -4}{8 d (d+2)} \Lambda^{-4}
\label{eddyvisc}
\eeq
has been introduced. $\nu_{eddy}$ reflects the effect of turbulence
acting on length scales smaller than $\Lambda^{-1}$ onto 
the geometric flow at scales larger than $\Lambda^{-1}$.
Curiously, note that $\nu_{eddy}$ is positive
for $d=3$ space dimensions, while \emph{it is negative for
$d=2$}. In other words, the effective viscosity in 
a system exhibiting two-dimensional
turbulence is smaller than the microscopic viscosity, due to the 
presence of a negative eddy viscosity. This interesting 
phenomenon has been known for decades, starting with the work of Starr 
and Kraichnan \cite{starr,kraich} (see also \cite{frisch,suko1}
for other approaches).

In heavy-ion collisions, one typically follows Bjorken in assuming
boost-invariance in the longitudinal direction \cite{Bjorken:1982qr}.
Therefore, the dynamics of relevance for fluid dynamics is effectively 
two-dimensional. Hence, if fluid turbulence develops, 
one expects the eddy viscosity to be negative and thus the 
fluid at RHIC would appear more ideal than it is based 
on its microscopic viscosity.

The physics of the appearance of a negative eddy viscosity
seem to be tied to the phenomenon of inverse energy cascade.
While in three dimensional turbulence, 
energy generally cascades down to smaller and 
smaller length scales until it is finally dissipated into heat
(regular cascade), in two dimension this process can 
seemingly reverse: energy is transferred from smaller to larger
length scales (see also the discussion in Ref.~\cite{kraich}).
%Simply put, small eddies convey energy to larger eddies and
%which 
A key ingredient seems to be that vorticity 
is conserved in two-dimensional incompressible fluids, allowing
small eddies to convey energy to larger eddies.
%such that
%eddies can convey energy towards larger length scales.
For compressible relativistic fluids, vorticity is no longer
conserved \cite{Romatschke:2007mq}, and hence one can expect the 
inverse cascade to ``leak'' energy. Indeed, working
out the turbulent correction to the sound mode of the propagator,
one finds an $r_B$ which is identical to $r_A$  except
that in (\ref{rA}) one has to replace
%\beq
%r_B=\alpha \nu^{-2} k^2 \frac{1}{(16\pi)} \left(\frac{7}{4}-4 c_s^2+c_s^4\right)
%\Lambda^{-4} + {\mathcal O} (k^3),
%\label{rB}
%\eeq
$$
\frac{d^2-d-4}{d(d+2)}\rightarrow \frac{2 (d^2-1)-2 c_s^2 (d-1)(d+2)}{
d(d+2)}+\frac{d (1-c_s^2)^2}{2 (d-1)^2}. 
$$
This implies that $r_B>0$ for both $d=2,3$ and $c_s^2\sim \frac{1}{3}$.

It seems that in two-dimensional compressible fluid turbulence
the shear mode gets less dissipative while for sound 
the converse is true. This means that it could be
difficult to decide whether to expect an increase
or decrease of effective viscosity in different experimental
observables.
%
%Different experimental observables
%can be more sensitive to one or the other, making 
%it difficult to decide whether to expect an increase
%or decrease in apparent viscosity. 
%However, 
%it may be that Morkovin's Hypothesis, wh
%
The very small apparent viscosity at RHIC has been extracted
mainly from the experimental observable called ``elliptic
flow'' \cite{Ollitrault:1992bk,Romatschke:2007mq}.
Invoking Morkovin's Hypothesis that under
certain conditions in compressible turbulence
``the essential dynamics of these shear flows will follow
the incompressible pattern'' \cite{smitdus},
in the following it will be speculated that 
elliptic flow is mostly influenced by the negative
eddy viscosity in the shear mode.
%the shear mode is dominating.
%
%In the following let us speculate
%that this observable indicates such a small
%$\eta/s$ because it is mostly influenced by the negative
%eddy viscosity in the shear mode. 
Then, is it possible
to verify or falsify the idea of fluid turbulence
in relativistic heavy-ion collisions?
%
%
%Both phenomena are then expected to
%occur at RHIC, but it 
%Obviously, it remains to be shown that fluid turbulence causing
%a negative eddy viscosity is indeed the reason for the small apparent
%viscosity seen at RHIC. In order to verify or falsify this idea,
%some other experimentally accessible prediction from Eq.(\ref{eddyvisc})
%is needed. 

Clearly, some other experimentally accessible 
prediction from Eq.~(\ref{eddyvisc}) is needed.
To do so, first introduce the dimensionless coupling parameter 
$\lambda=\frac{\alpha}{\nu^3 \Lambda^4}$ so that $\nu_{eff}=\nu (1-\frac{\lambda}{64 \pi})$,
where $d=2$ has been used in Eq.~(\ref{eddyvisc}).
%is a positive number that can be read off from 
Now a renormalization group improvement is performed by solving
the set of equations \cite{YO}
\beq
\frac{d \nu_{eff}}{d \Lambda}=\nu_{eff} \frac{\lambda}{16 \pi \Lambda},\qquad
\lambda=\frac{\alpha}{\nu_{eff}^3 \Lambda^4},
\eeq
where to leading order in $\lambda$, 
$\nu$ has been replaced by $\nu_{eff}$ and $\alpha$ was assumed to be independent
of $\Lambda$ \cite{YO}. Using $\nu_{eff}|_{\lambda=0}=\nu$, the result is
$\nu_{eff}=\nu \left(1-\frac{3\lambda}{64 \pi}\right)^{1/3}$.
In other words, one expects the effective ratio $\eta/s$ to behave as
\beq
\left(\frac{\eta}{s}\right)_{eff}=\left(\frac{\eta}{s}\right)
\left[1-\bar{\lambda} \left(\frac{\eta}{s}\right)^{-3}\right]^{1/3},
\label{effetas}
\eeq
where for convenience $\bar{\lambda}=\frac{3\alpha T^3}{64 \pi \Lambda^4}$ 
has been introduced.
%is also a dimensionless parameter. 
It may be that $\bar{\lambda}$ still
depends on $\eta/s$:  for the following qualitative discussion, 
this should not matter unless $\bar{\lambda}$ is proportional to 
$\eta/s$ with a power greater or equal three. $\bar{\lambda}$ will
typically be rather small: since $\alpha$ has to be proportional to 
the temperature (the only other dimensionful scale except $\Lambda$),
one expects $\bar{\lambda}\sim (T/\Lambda)^4$. Assuming
$\Lambda$ corresponds to the smallest resolved length scale of
current %viscous hydrodynamic simulations 
heavy-ion collision simulations
\cite{Romatschke:2007mq}
one has $\Lambda\sim 1$ GeV, so using again $T\sim 200$ MeV gives
$\bar{\lambda}\sim 10^{-3}$.

%\footnote{$\bar{\lambda}$
%may turn out to depend on $\eta/s$. For the qualitative discussion, 
%this should not matter unless $\bar{\lambda}$ is proportional to 
%$\eta/s$ with a power greater or equal three. $\bar{\lambda}$ will

%}. 

\begin{figure}
%\vspace{0.6cm}
\begin{center}
\includegraphics[height=5cm]{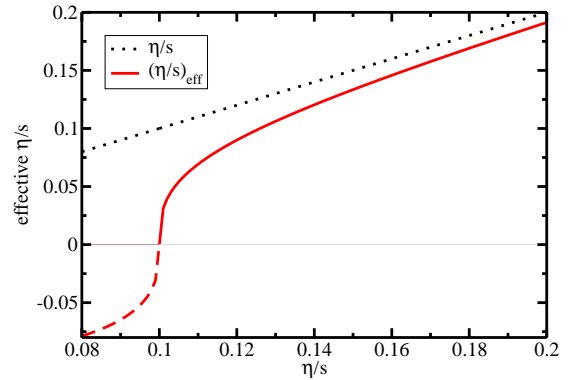}
\end{center}
\caption{Effective versus standard viscosity for $\bar{\lambda}=10^{-3}$. 
Decreasing $\eta/s$ from large values, the effective viscosity changes rapidly 
to extremely small values close to the critical
viscosity $\eta/s=(\bar{\lambda})^{1/3}$. Decreasing $\eta/s$ further
may even result in negative effective viscosity. However, at this point
the calculation breaks down and the result cannot be trusted.}
\label{fig1}
\end{figure}

The qualitative behavior of the effective viscosity is shown in
Fig.~\ref{fig1} for $\bar{\lambda}=10^{-3}$. It can be seen that close 
to the critical viscosity the effective viscosity becomes very small
and there is some indication that even slightly negative values of
effective viscosities can be achieved. It is expected that
higher non-linearities subsequently render the effective viscosity 
again positive \cite{frisch}. Since these terms have been ignored 
in the above calculation, the result in Eq.~(\ref{effetas}) cannot
be trusted below the critical viscosity.
Nevertheless, the strong suppression of the effective viscosity close 
to the critical viscosity suggested by Eq.~(\ref{effetas}) 
should be experimentally accessible,
e.g. by a measurement of the elliptic flow 
at the upcoming
Large Hadron Collider (LHC). The magnitude of elliptic flow is known
to decrease for increasing $\eta/s$ \cite{Teaney:2003kp}.
Turning the argument around, if fluid turbulence is in operation at RHIC,
and elliptic flow is indeed mostly affected by the negative eddy viscosity
in the shear mode, the effect at the LHC should be even more 
pronounced: from Fig.\ref{fig1}
one would expect the elliptic flow to increase beyond the RHIC values,
maybe even beyond the ``ideal hydrodynamic limit'', if a negative
effective viscosity is realized for an extended period of time.
Note that more conventional approaches \cite{Krieg:2007sx,Molnar:2007an} 
call for a decrease of elliptic flow at the LHC, while 
extrapolations from existing data do indicate an increase
\cite{Borghini:2007ub}.

It should be stressed that the above calculation of the eddy viscosity for 
fluid turbulence in relativistic heavy-ion collisions 
can be regarded as qualitative at best. However, given interest
in the community, there are many ways to improve or strengthen
%on %the %qualitative result 
Eq.~(\ref{effetas}), e.g. by relaxing various assumptions 
made in the derivation or building upon the considerable knowledge
%
%recent advances
from studying two-dimensional incompressible fluid turbulence 
(see e.g. \cite{2dNRstud} and references therein).
%
%. Maybe most importantly, the value
%of $\bar{\lambda}$ and thus the critical viscosity 
%could be calculated by 
%deriving a relation between the fluctuation 
%strength $\alpha$ and $\nu,\epsilon$. Such relations are known
%for systems in equilibrium \cite{mori}. 
%\textbf{3 point functions? Mike's telling apart hydro visc from aniso?}

Finally, it should be pointed out that the concept of anomalous
turbulent viscosity in the context of heavy-ion collisions 
has been already suggested in Ref.~\cite{Asakawa:2006tc}. However, the
discussion in \cite{Asakawa:2006tc} is based on assuming the presence 
of plasma turbulence, which is somewhat different than the
fluid turbulence picture outline here. (Non-Abelian) Plasma turbulence
seems to occur as a consequence of plasma instabilities 
\cite{Mrowczynski:2005ki,Arnold:2005ef,Dumitru:2006pz,Romatschke:2006nk}.
However, given current
estimates of initial plasma parameters 
%(such as initialization time
%and density), 
it seems somewhat unlikely that plasma instabilities
set in early enough  \cite{Romatschke:2006nk,Romatschke:2006wg}
for plasma turbulence to be relevant at RHIC. 
%(``too little, too late'') 
Nevertheless, the spectrum of initial fluctuations studied in
this context \cite{Fukushima:2006ax} 
could also be of relevance for fluid turbulence.

To summarize, following the Yakhot-Orszag approach to turbulence
the eddy viscosity in a compressible fluid with a relativistic
equation of state was calculated. The result differs 
from incompressible fluids by the presence of a sound channel
in the fluid dynamic propagator. 
In the sound channel, the eddy viscosity is positive for both
two and three dimensions. 
In the shear channel, the
eddy viscosity is found to be identical to that of incompressible fluids,
and happens to be negative for two space dimensions.
Given the Reynolds number for RHIC experiments is expected to be
much larger than unity, and that the relevant fluid dynamics is
essentially two-dimensional, it may be that this explains
the apparently tiny viscosity over entropy density ratio
extracted from experimental data. If this is the case, the model
would predict a substantially increased elliptic flow at the LHC,
which is experimentally falsifiable in the near future.

\acknowledgments

I would like to thank K.~Kajantie, J.~Riley, P.~Rhines, D.T.~Son and V.~Yakhot 
for fruitful discussions.
This work was supported by the US Department of Energy, grant
number DE-FG02-00ER41132.

\end{document}